\documentclass[adp,fleqn]{w-art}
\usepackage{times}
\usepackage{w-thm}
\usepackage[]{graphicx}
\chardef\bslash=`\\ 

\newcommand{\nn}{\nonumber}
\newcommand{\be}{\begin{equation}}
\newcommand{\ee}{\end{equation}}
\newcommand{\ba}{\begin{eqnarray}}
\newcommand{\ea}{\end{eqnarray}}

\def\gev{~{\rm GeV}}

\newcommand{\lsim}{\raisebox{-4pt}{$\,\stackrel{\textstyle
                                                         <}{\sim}\,$}}

\newcommand{\req}[1]{(\ref{#1})}

\def\sh{{s}}
\def\uh{{u}}
\def\={\,=\,}

\begin{document}
\pagespan{1}{}



\title[handbag]{Some comments on the handbag approach
  to wide-angle exclusive scattering}

\author[F.\ Kroll]{P.\ Kroll\footnote{
     E-mail: {\sf kroll@physik.uni-wuppertal.de}, Phone: +49\,202\,439\,2620,
     Fax: +49\,202\,439\,3860}\inst{1}} 
\address[\inst{1}]{Fachbereich Physik, Universit\"at Wuppertal,\\
D-42097 Wuppertal, Germany }
\begin{abstract}
The handbag mechanism for wide-angle exclusive scattering reactions is 
discussed and compared to other theoretical approaches. The role of
power laws in observables is critically examined. Applications of the
handbag mechanism to Compton scattering and meson photoproduction are
presented. The soft physics input to these processes are specific form
factors which represent $1/x$ moments of generalized parton
distributions at zero skewness. A recent analysis of the nucleon form
factors provides these GPDs and, hence, the new form factors. 
\end{abstract}
\maketitle                   

\section{Introduction}
\label{sect:intro}
\begin{figure}[t]
\begin{center}
\includegraphics[width=4.6cm,bbllx=120pt,bblly=570pt,bburx=265pt,
bbury=680pt,clip=true]{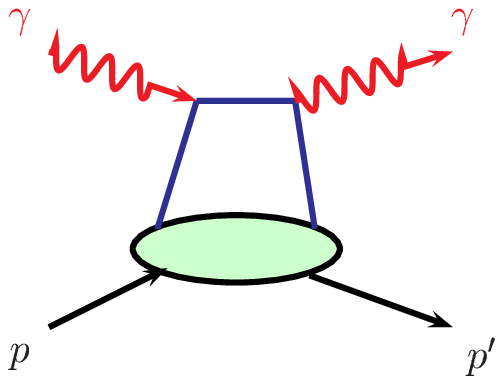} \hspace*{0.3cm}
\includegraphics[width=4.6cm,bbllx=340pt,bblly=580pt,bburx=485pt,
bbury=693pt,clip=true]{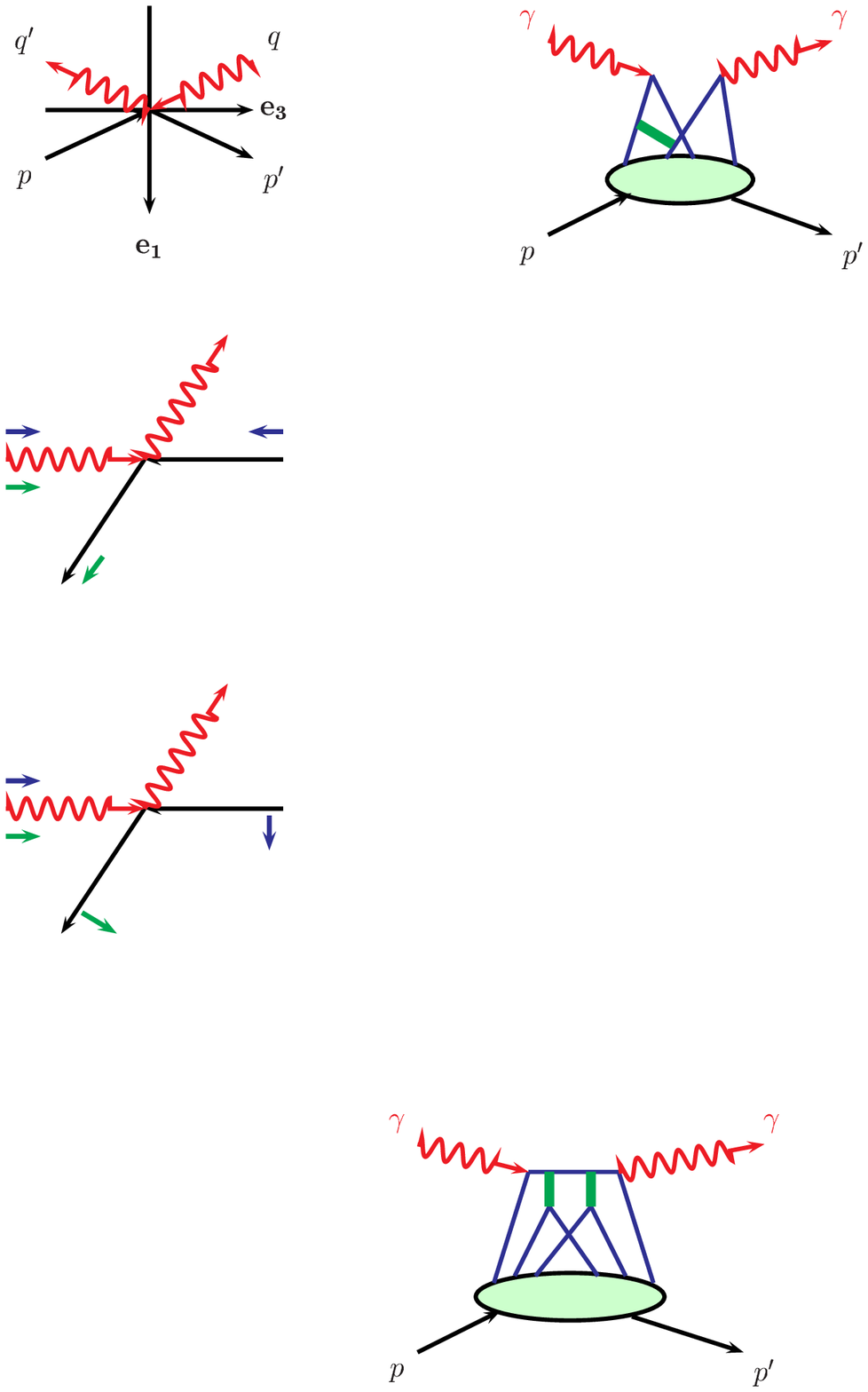}\\[0.2em] 
\includegraphics[width=4.6cm,bbllx=285pt,bblly=125pt,bburx=450pt,
bbury=240pt,clip=true]{hand-graph.ps} \vspace*{-0.3cm} 
\includegraphics[width=5.0cm,bbllx=105pt,bblly=465pt,bburx=385pt,
bbury=610pt,clip=true]{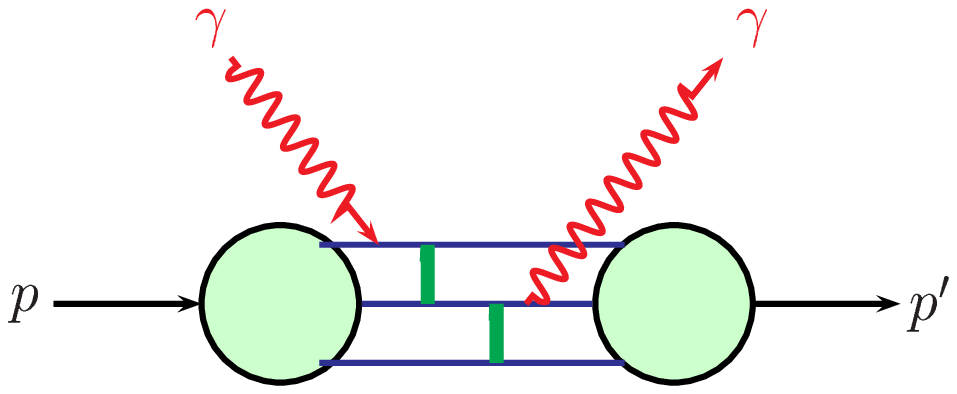} 
\caption{{}Handbag diagram for Compton scattering (upper left), cat's 
ears (upper right), the three-particle contribution (lower left) and
its valence quark approximation (lower left).}
\label{fig:handbag}
\end{center}
\end{figure}

Recently a new approach to wide-angle Compton scattering off protons
has been proposed \cite{rad98,DFJK1} where, for Mandelstam variables 
$s,-t,-u$ that are large as compared to the square of a typical
hadronic scale $\Lambda$ (being of  the order of $1\gev^2$), the
process amplitudes factorize into a hard parton-level subprocess, 
Compton scattering off quarks, and in soft form factors which
represent $1/x$ moments of generalized parton distributions 
(GPDs) and encode the soft physics (see Fig.\ \ref{fig:handbag}). 
Subsequently it has been realized that this so-called handbag mechanism 
also applies to a number of other wide-angle reactions such as 
meson photo- and electroproduction \cite{hanwen} or two-photon
annihilations into pairs of mesons \cite{DKV2} or baryons 
\cite{DKV2,weiss}. It should be noted that the handbag mechanism bears
resemblance to the treatment of inelastic Compton scattering advocated
for by Bjorken and Paschos \cite{bjo} long time ago.   

There are competing mechanisms which also contribute to wide-angle
scattering besides the handbag which is characterized by one active
parton, i.e.\ one parton from each hadron participates in the hard 
subprocess (e.g.\ $\gamma q\to \gamma q$ in Compton scattering) while
all others are spectators. First there are the so-called cat's ears 
graphs (see Fig.\ \ref{fig:handbag}) with two active partons 
participating in the subprocess (e.g.\ $\gamma qq \to \gamma qq$). 
It can be shown that in these graphs either a large parton virtuality 
or a large parton transverse momentum occurs. This forces the exchange 
of at least one hard gluon in the subprocess. Hence, the cat's ears 
contribution is expected to be suppressed as compared to the handbag
one. The soft physics in the case of the cat's ears are hadronic
matrix elements that describe higher order quark correlations inside
the proton. Nothing is known about these matrix elements as yet. 

The next class of graphs is characterized by three active quarks and,
obviously, requires the exchange of at least two hard gluons. One can go on
and consider four or more active partons. This way one obtains an
expansion of the scattering amplitude which bears resemblance to the
series of contributions from n-body operators appearing in
non-relativistic many-body theory. In principle, all the different 
contributions have to be added coherently. In practice, however, this 
is a difficult, currently almost impossible task since each
contribution has its own associated soft hadronic matrix elements
which, as yet, cannot be calculated from QCD and are often even 
phenomenologically unknown. 

\section{Comments on the leading-twist contribution}
\label{sec:comments}
At large momentum transfer the restriction to valence quarks is a
reliable approximation to exclusive scattering. Exploiting it in the
case of three active quarks, the big blob~\footnote{This holds for 
 protons. In the case of, say, Compton scattering off pions the cat's 
 ears already include the leading-twist contribution.} 
decays into two smaller blobs, see Fig.\ \ref{fig:handbag}. These
blobs describe the proton's distribution amplitude for finding valence 
quarks in the hadron, each carrying some fraction $x_i$ of the
hadron's momentum.  This contribution is the so-called leading-twist 
contribution \cite{bro80} which is expected to dominate for
asymptotically large momentum transfer but seems to be way below
experiment for momentum transfer of the order of $10\, \gev^2$. Actual
calculations of Compton scattering \cite{dixon} and the proton form
factor \cite{bolz} confirm this assertion. There are only very few exceptions 
where the leading-twist contribution is close to experiment. One of
those is the $\pi-\gamma$ transition form factor. This is a special
case since here the handbag with one active quark and the
leading-twist approximation - for which all valence quarks participate
in the hard subprocess - fall together. Indeed there is general
agreement in the literature that the leading-twist/handbag mechanism
provides the bulk of the contribution to that form factor. The other
exceptions are some exclusive charmonium decays into light hadrons
where something like a (time-like version of the) handbag does not occur 
unless one allows for intrinsic charm in the light hadrons. Thus, 
for instance, the $J/\psi$ decay into a baryon-antibaryon pair is 
dominated by the leading-twist contribution where the $c\bar{c}$ pair from the 
$J/\psi$ decay annihilates into three gluons which subsequently turn
into light $q\bar{q}$ pairs which form the final state hadrons. As has
been shown \cite{bolz96} this contribution is large enough to account 
for the measured decay width. I have however to remind the reader that 
there is a number of exclusive charmonium decays, characterized by 
non-conservation of hadronic helicity, which are not under control of 
the leading-twist mechanism. An example is set by the process 
$J/\psi\to \rho\pi$, the famous $\rho\pi$ puzzle.

A feature of the leading-twist contribution is the power behaviour of
form factors and cross sections \cite{bro-far}. These observables
decrease by powers of the hard scale asymptotically. The powers are
determined by the number of partons participating in the hard process
\footnote{The asymptotic power laws hold if the hard scale is much
  larger than any soft scale available in the process (e.g.\ hadronic
  masses). In almost all data on exclusive reactions this is however
  not the case.}. 
It is to be stressed that the powers of the hard scale are
accompanied by perturbative logs for which there is no experimental  
evidence in exclusive wide-angle scattering as yet in contrast to deep 
inclusive lepton-nucleon scattering where they are seen in experiment.
Their role in the extraction of the parton distribution from the
inclusive lepton-nucleon data cannot be overemphasized. Thus, an 
approximate power behaviour observed in an exclusive observable in a
finite and typically rather limited range of the hard scale cannot be
considered as evidence for the dominance of the leading-twist mechanism. 
This a premature: a careful analysis of the normalization and the role of the
perturbative logs is mandatory. Furthermore an estimate of the size of 
contributions from alternative mechanisms like the handbag one, is
required before this conclusion can be drawn.

\begin{figure}
\begin{center}
\includegraphics[width=.47\textwidth,
  bb=100 355 470 705,clip=true]{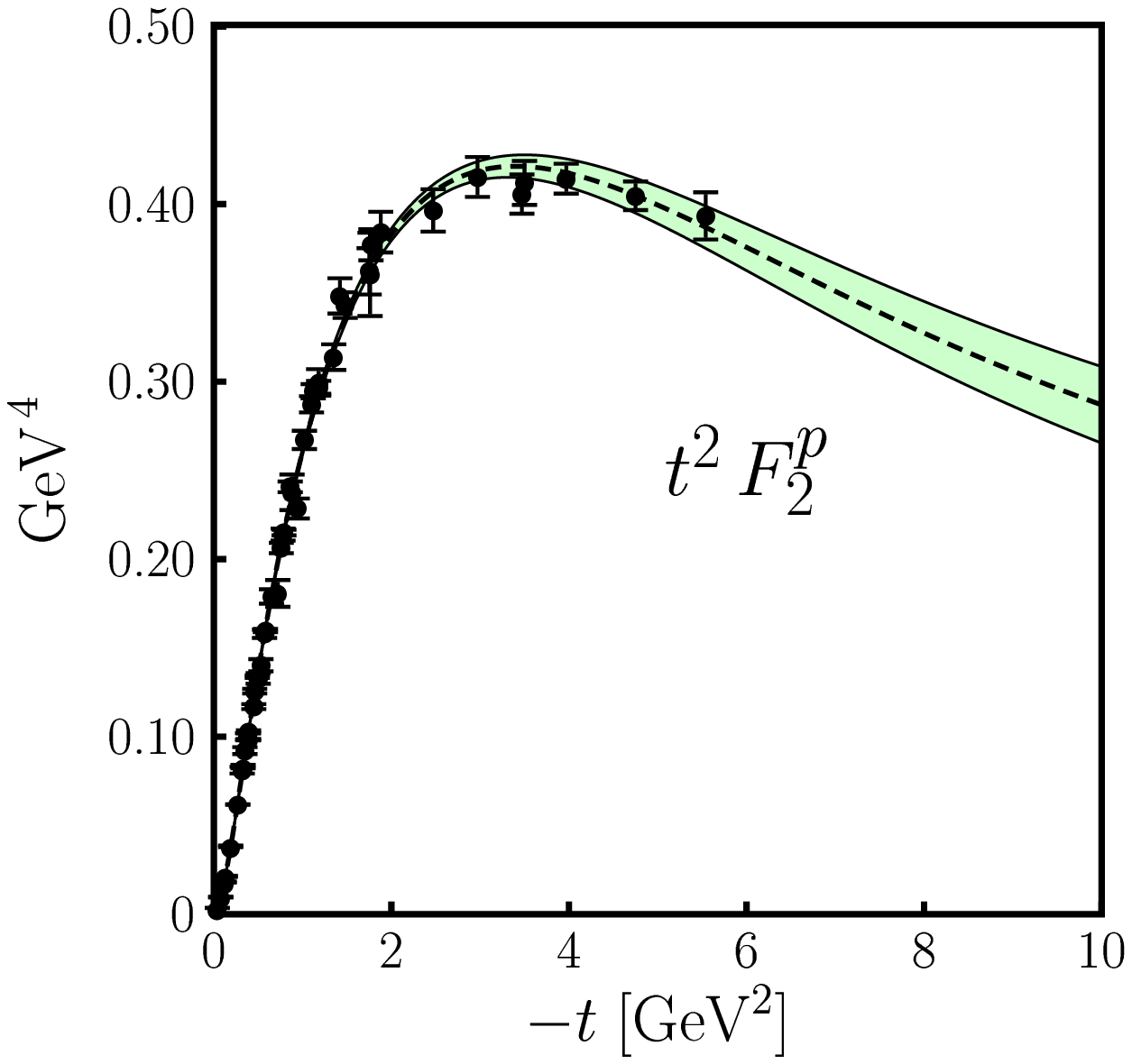}
 \end{center}
\caption{\label{fig:F2data} Large t data \cite{gayou} for the scaled
Pauli form factor $t^2 F_2^p(t)$ of the proton (data at lower t  are
also shown). The dashed line together with the $1\sigma$ error band
represents the phenomenological results obtained in \protect\cite{DFJK4}.} 
\end{figure}

An interesting example is set by the recent measurement of the ratio
of the proton's Pauli and Dirac form factor \cite{gayou}. The
observation that $\sqrt{-t}F_2/F_1$ exhibits a nearly constant
behaviour for the largest measured momentum transfers ($2-6\,\gev^2$),  
stirred a hot debate on the asymptotic behaviour of $F_2$. It is
however more elucidating to isolate $F_2$ from the data and to look 
to its behaviour instead to the ratio $\sqrt{-t}F_2/F_1$. In Fig.\
\ref{fig:F2data} the form factor data, scaled by $t^2$, are displayed. The
data rather show a dipole behaviour in that range of $t$ than the
asymptotically expected $\sim t^{-3}$ fall off. Nothing is wrong with
that result. The range $2\,\gev^2 \lsim -t \lsim 6\,\gev^2$ is simply
too small for probing the asymptotic behaviour. Soft physics still dominates,
that is all.      

Soft mechanisms like the handbag can easily produce a power behaviour
of form factors. Consider for instance the following ansatz for the
GPD $H$ at zero skewness and for valence quarks 
\be
H^q_v(x,t)\= q_v(x) \exp{[tf_q(x)]}\,,
\label{ansatz}
\ee 
which is advocated for in \cite{DFJK4}. Here, $q_v$ is the usual
parton distribution for valence quarks. As shown in  \cite{DFJK4}, at
large $t$ the dominant contribution to the Dirac form factor, $F_1$,
comes from a rather narrow region of large $x$. Thus, one may take a
large-$x$ approximations of \req{ansatz}: $q_v\sim (1-x)^{\beta_q}$ and 
$f_q\sim A_q (1-x)^2$ where $A_q$ is a soft parameter of the order of 
$1\,\gev^{-2}$. The power $\beta_q$ is provided by the phenomenological 
parton distributions \cite{Cteq}: $\beta_u\simeq 3.4$, $\beta_d\simeq 5$. 
The function \req{ansatz} has a maximum at 
\be 
1-x_s \= \sqrt{\frac{\beta_q}{2A_q|t|}}\,.
\label{saddle}
\ee
Hence, the sum rule 
\be
h^q_{1,0}(t)\=\int_0^1 dx H_v^q(x,t)\,,
\ee
can be evaluated in the saddle point approximation and one finds the
power law
\be
h^q_{1,0} \sim |t|^{-(1+\beta_q)/2}\,.
\label{powers}
\ee
With $\beta_u\simeq 3.4$ we see that the $u$-quark contribution to the
Dirac from factor, $h^u_{1,0}$, falls slightly faster then $t^{-2}$ while the
$d$-quark contribution drops as $t^{-3}$. It is to be stressed that
this is not an asymptotic result but holds provided the saddle point
\req{saddle} lies in region in which the bulk of the contribution to
the Dirac form factor is accumulated. To this region charaterized by 
$1-x \sim \Lambda/\sqrt{|t|}$ ($\Lambda$ is a typical hadronic scale of
order $1\,\gev$), the Feynman mechanism applies for which the
struck quark carries most of the proton momentum and thus large
internal virtualities of order $t$ are avoided. As shown in \cite{DFJK4} the
ansatz \req{ansatz} works quite well for $t$ values up to
$30\,\gev^2$. For very large values of $t$ one may expect the soft
contribution to be damped by Sudakov factors and the leading-twist 
contribution may take the lead.  
\section{Handbag factorization for wide-angle Compton scattering}
\label{sec:wacs}
Consider Mandelstam variables $s$, $-t$ and $-u$ that are large as
compared to $\Lambda^2$. The contribution from the handbag diagram shown in
Fig.\ \ref{fig:handbag}, is calculated in a symmetrical frame which is
a c.m.s.\ rotated in such a way that the momenta of the incoming ($p$) 
and outgoing ($p'$) proton momenta have the same light-cone plus 
components. Hence, the skewness defined as 
\be 
\xi \= \frac{(p - p')^+}{(p + p')^+}\,,
\ee
is zero. The bubble in the handbag is viewed as a sum over all possible 
parton configurations as in deep ineleastic lepton-proton scattering. 
The crucial assumption in the handbag approach is that of restricted
parton virtualities, $k_i^2<\Lambda^2$, and of intrinsic transverse
parton momenta, ${\bf k_{\perp i}}$, defined with respect to their
parent hadron's momentum, which satisfy $k_{\perp i}^2/x_i
<\Lambda^2$, where $x_i$ is the momentum  fraction parton $i$ carries.   
 
One can then show \cite{DFJK1} that the subprocess Mandelstam variables
$\hat{s}$ and $\hat{u}$ are the same as the ones for the full process,
Compton scattering off protons, up to corrections of order
$\Lambda^2/t$:
\ba
\hat{s} &=& (k_j+q)^2 \simeq (p+q)^2 \=s\,, \nn\\ 
\hat{u} &=& (k_j-q')^2 \simeq (p-q')^2 \=u\,.
\ea
The active partons, i.e.\ the ones to which the photons couple, are
approximately on-shell, move collinear with their parent hadrons and
carry a momentum fraction close to unity. Thus, like in deep virtual 
Compton scattering, the physical situation is that of a hard 
parton-level subprocess, $\gamma q\to \gamma q$, and a soft emission 
and reabsorption of quarks from the proton. 
 
The light-cone helicity amplitudes \cite{diehl01} for 
wide-angle Compton scattering then read \cite{DFJK1,HKM}
\ba
{M}_{\mu'\nu',\,\mu \nu}(s,t) &=& \; \frac{e^2}{2}\,
             \Big[\, \delta_{\nu'\nu}\,{ T}_{\mu'\nu,\,\mu\nu}(\sh,t)\, 
                             (R_V(t) + R_A(t))  \nn\\[0.3em]
 &+& \, \delta_{\nu'\nu}\,{ T}_{\mu'-\nu,\,\mu -\nu}(\sh,t)\, 
                        (R_V(t) - R_A(t)) \nn\\[0.3em] 
  &+& \,\delta_{-\nu'\nu}\,\frac{\sqrt{-t}}{2m} 
         \left(\,  T_{\mu'-\nu',\,\mu\nu}(\sh,t)\, 
   + { T}_{\mu'\nu',\,\mu -\nu}(\sh,t)\, \right) \Big] \,R_T(t)\,. 
\label{ampl}
\ea
The labels $\mu (\nu),\, \mu'(\nu')$ denote the helicities of the
incoming and outgoing photons (protons in $M$ and quarks in the
subprocess amplitude $T$), respectively. The mass of the proton 
is denoted by $m$. The Compton form factors $R_i$  represent $1/x$-moments 
of GPDs at zero skewness. This representation which requires the
dominance of the plus components of the proton matrix elements, is a
non-trivial feature given that, in contrast to deep inelastic 
lepton-nucleon and deep virtual Compton scattering, not only the plus 
components of the proton momenta but also their minus and transverse 
components are large here. 
For Compton scattering the hard scattering has been calculated to 
next-to-leading order (NLO) perturbative QCD \cite{HKM}. The infrared
singularities occuring to this order can be absorbed into the Compton
form factors. To NLO one also has to take into account the 
photon-gluon subprocess and a corresponding gluonic form factor for 
consistency. This small correction which amounts to less than $10\%$ 
in cross section, is taken into in the numerical results shown below
but, for convenience, ignored in all following formulas.  

The handbag amplitudes \req{ampl} lead to the following leading-order 
result for the Compton cross section  
\ba
\frac{d\sigma}{dt} &=& \frac{d\hat{\sigma}}{dt} \left\{ \frac12\, \big[
R_V^2(t)\, + \frac{-t}{4m^2} R_T^2(t) + R_A^2(t)\big] \right.\nn\\
&&\hspace*{-0.5cm}\left.  - \frac{\uh\sh}{\sh^2+\uh^2}\,
\big[R_V^2(t)\,+ \frac{-t}{4m^2} R_T^2(t) - R_A^2(t)\big]\right\}\,,
\label{dsdt}
\ea
where $d\hat{\sigma}/dt$ is the Klein-Nishina cross section for
Compton scattering off massless, point-like spin-1/2 particles of
charge unity. The NLO corrections are not shown in \req{dsdt}.

Another interesting observable in Compton scattering is the helicity
correlation, $A_{LL}$,  between the initial state photon and proton
or, equivalently, the helicity transfer, $K_{LL}$, from the incoming
photon to the outgoing proton. In the handbag approach one obtains
\cite{HKM,DFJK2} 
\be
A_{LL}\=K_{LL}\simeq \frac{\sh^2 - \uh^2}{\sh^2 + \uh^2}\, 
                    \frac{R_A(t)}{R_V(t)} + O(\alpha_s)\,,
\label{all}
\ee  
where the factor in front of the form factors is the corresponding
observable for $\gamma q\to \gamma q$. The result \req{all} is a
robust prediction of the handbag mechanism, the magnitude of the
subprocess helicity correlation is only diluted  somewhat by the 
ratio of the form factors $R_A$ and $R_V$. It is to be stressed that
$A_{LL}$ and $K_{LL}$ coincide in the handbag approach
because the quarks are assumed to be massless and hence there is no
quark helicity flip. For an alternative approach, see Ref.\ \cite{miller}. 
\section{A phenomenological analyis of the GPDs}
\label{sect:analysis}
In order to make actual predictions for Compton scattering  
the form factors or rather the underlying GPDs are  required.
A first attempt to determine the GPDs in analogy to the analyses of
the usual parton distributions has been performed recently
\cite{DFJK4}. A parameterization of the GPDs
is used in that analyses that interpolates between the expected Regge
behaviour at small $t$ and very small $x$  and an overlap model that
applies to large $t$ and large $x$ \cite{rad98,DFJK1,DFJK3}. The
ansatz used in \cite{DFJK4} is already given in \req{ansatz}. The
profile function is assumed to read
\be
f_q(x)\= [\alpha' \log(1/x) + B_q](1-x)^{n+1} + A_qx(1-x)^n\,,
\label{exponent}
\ee 
with $n=1,2$ and $\alpha'=0.9\,\gev^{-2}$ is the usual Regge slope. In 
an analogous way the GPDs $\widetilde{H}$ and $E$ are parameterized 
with the additional complication that the forward limit of $E$ is not 
accessible in deep inelastic lepton-nucleon scattering and, hence, is 
to be determined in that analysis too. The free parameters of this 
approach are fitted to the available data on the nucleon form factors 
$F_{1,2}^{p,n}$ and $F_A$ at all $t$. The fit to the form factors
through the sum rules 
\be
F_1^{p(n)}(t) \= e_{u(d)} \int_0^1 dx H^u_v(x,t) 
                         + e_{d(u)} \int_0^1 dx H^d_v(x,t)\,, 
\ee
and analogously for the other form factors, only allows an access to the 
valence quarks at zero skewness. An example of the fit to the data is
given in Fig.\ \ref{fig:F2data}. It turns out in the analysis presented in
\cite{DFJK4} that the contributions from a rather limited range of $x$
dominate the form factors. The range is shifted towards larger
$x$-values with increasing $-t$. This property of the parameterization
\req{ansatz}, \req{exponent} leads to the power behaviour discussed in
Sect.\ \ref{sec:comments} 
\begin{figure}[t]
\begin{center}
\includegraphics[width=0.45\textwidth,
  bb=60 320 410 565,clip=true]{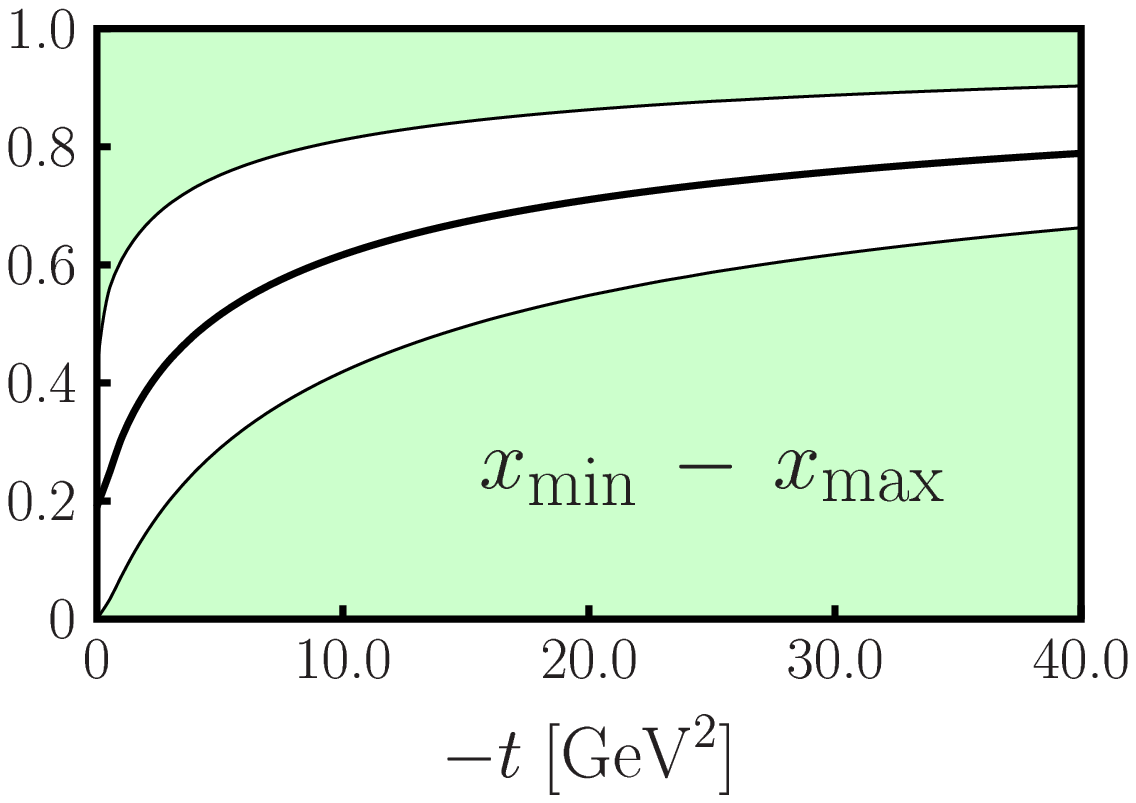}
\end{center}
\caption{\label{fig:var1_xminmax} Region of $x$ (white region) which
accounts for $90\%$ of $F_1^p(t)$ in the best fit to
(\protect\ref{ansatz}) at a scale $\mu=2\gev$.  The upper and lower shaded
$x$-regions each account for $5\%$ of $F_1^p(t)$.
The thick line shows the average $\langle x\rangle_t$.}
\end{figure}

The quality of the fits is very similar in the two
cases $n=1$ and 2 and the results for GPDs and related quantities agree very
well with each other. Substantial differences between the two results only occur
for $x$-values outside the range which is sensitive to the form
factor data (see Fig.\ \ref{fig:var1_xminmax}). It is the physical
interpretation of the results which favours the fit with $n=2$. Indeed
the average distance between the struck quark and the cluster of spectators
becomes unphysical large for $x\to \infty$ in the case of $n=1$.

\begin{figure}
\begin{center}
\includegraphics[width=.35\textwidth, height=.35\textwidth,
  bb=77 458 399 786,clip=true] {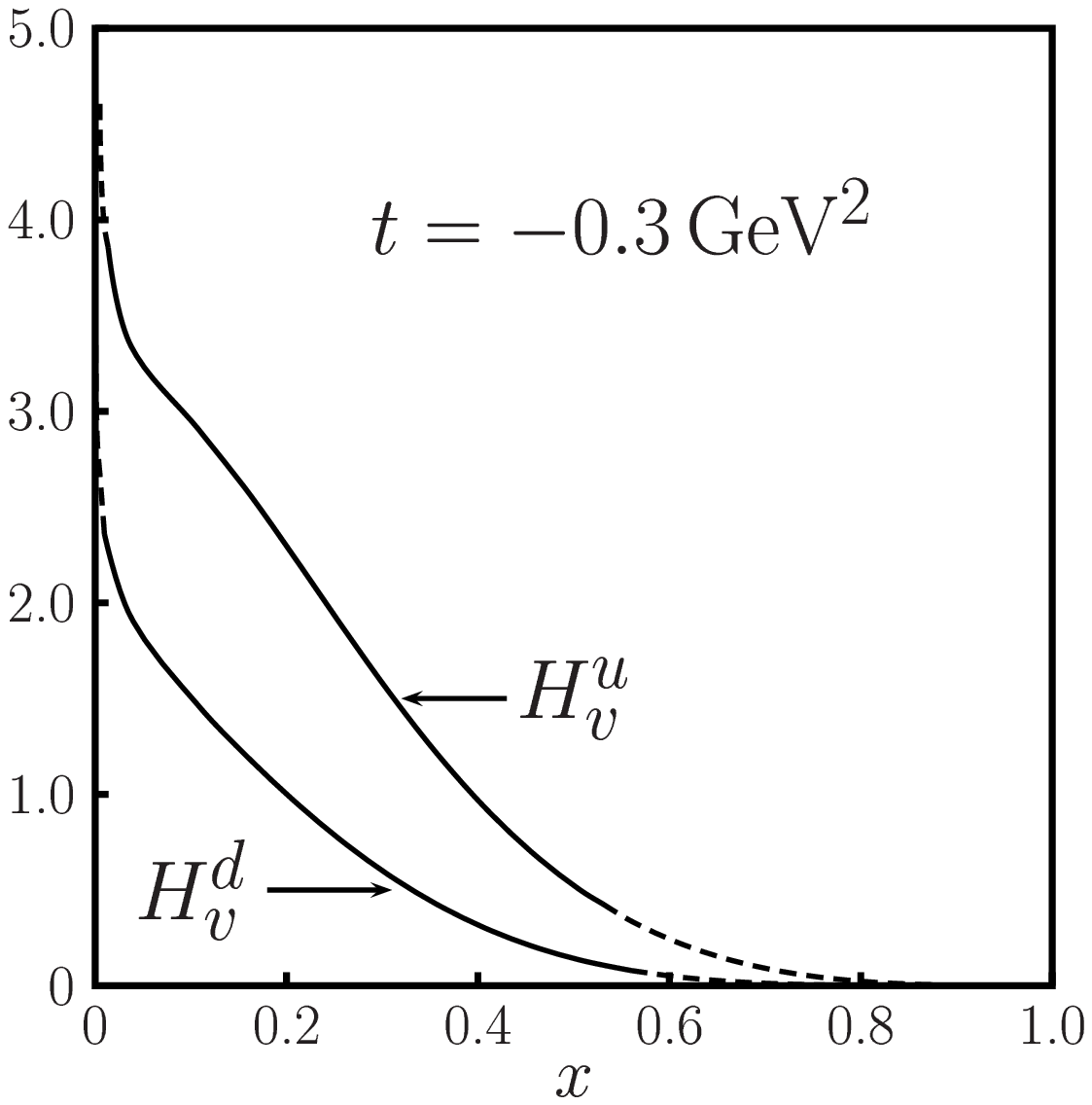}
\hspace{4em} 
\includegraphics[width=.35\textwidth, height=.35\textwidth,
  bb=68 364 399 683,clip=true] {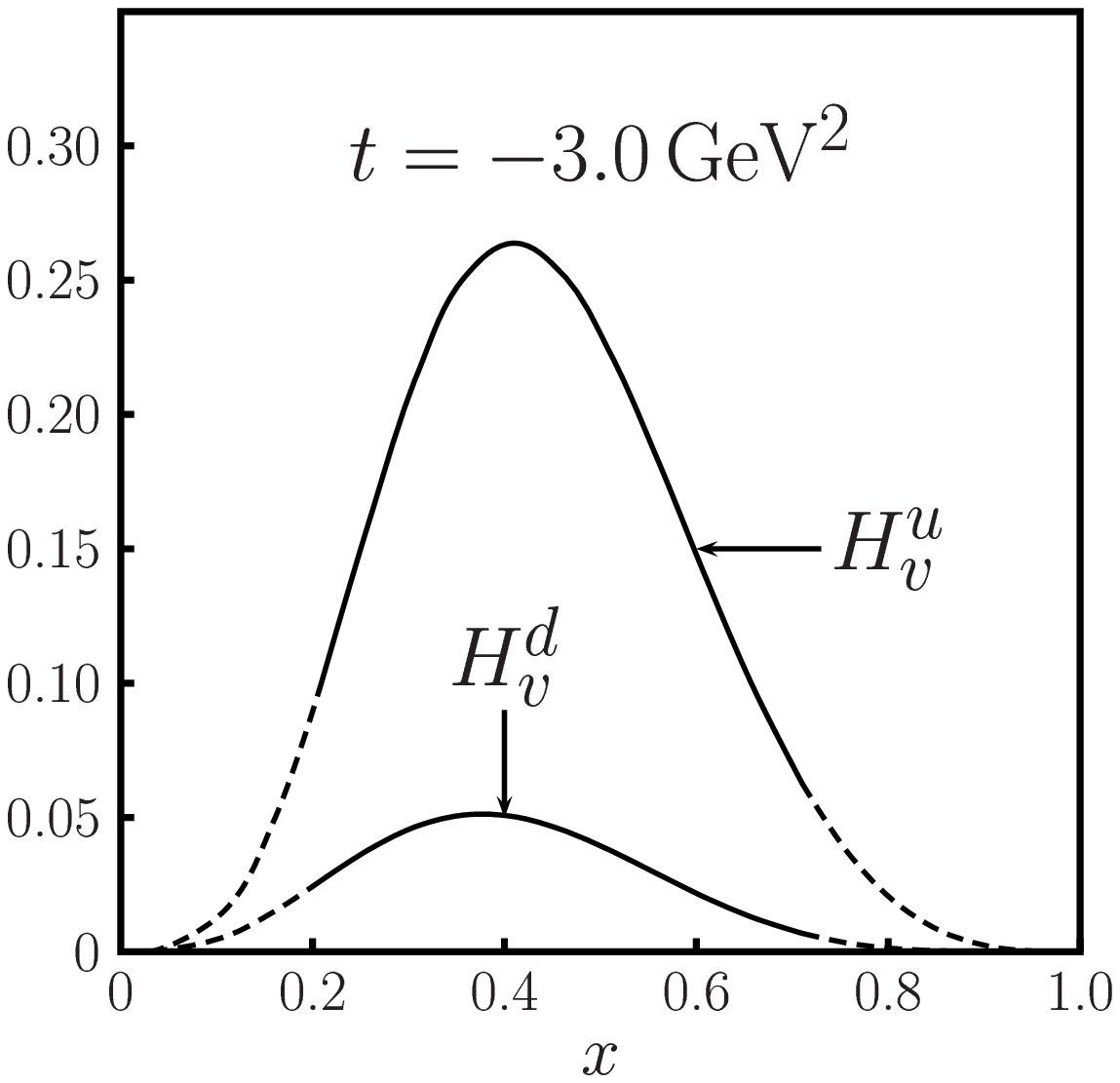}
\end{center}
\caption{\label{fig:Hgpd} Result for the valence GPDs $H_v^q(x,t)$ at
$\mu=2 \gev$ obtained in the analysis presented in \cite{DFJK4}.
Dashed lines indicate the regions where $x < x_{\rm min}(t)$ or $x >
x_{\rm max}(t)$.}
\end{figure}

As an example of the results for the GPDs $H$ is shown in Fig.\ \ref{fig:Hgpd}
at two values of $t$. While at small $t$ the GPD still
reflects the behaviour of the parton distribution it exhibits a
pronounced maximum at larger $t$. This maximum moves towards $x=1$
with increasing $-t$ and reflects the repeatedly mentioned property
that only a limited range of $x$ contributes to the form factors
substantially. The results for the GPDs $\widetilde{H}$ and $E$ behave
similarly. Noteworthy differences are that $\widetilde{H}$ and $E$ for
$u$ and $d$ quarks have opposite signs and that $E^u_v$ and $E^d_v$
are of about the same magnitude at least for smaller values of $t$.  

\begin{figure}
\begin{center}
\includegraphics[height=.35\textwidth,
 bb = 48 301 444 655,clip=true]{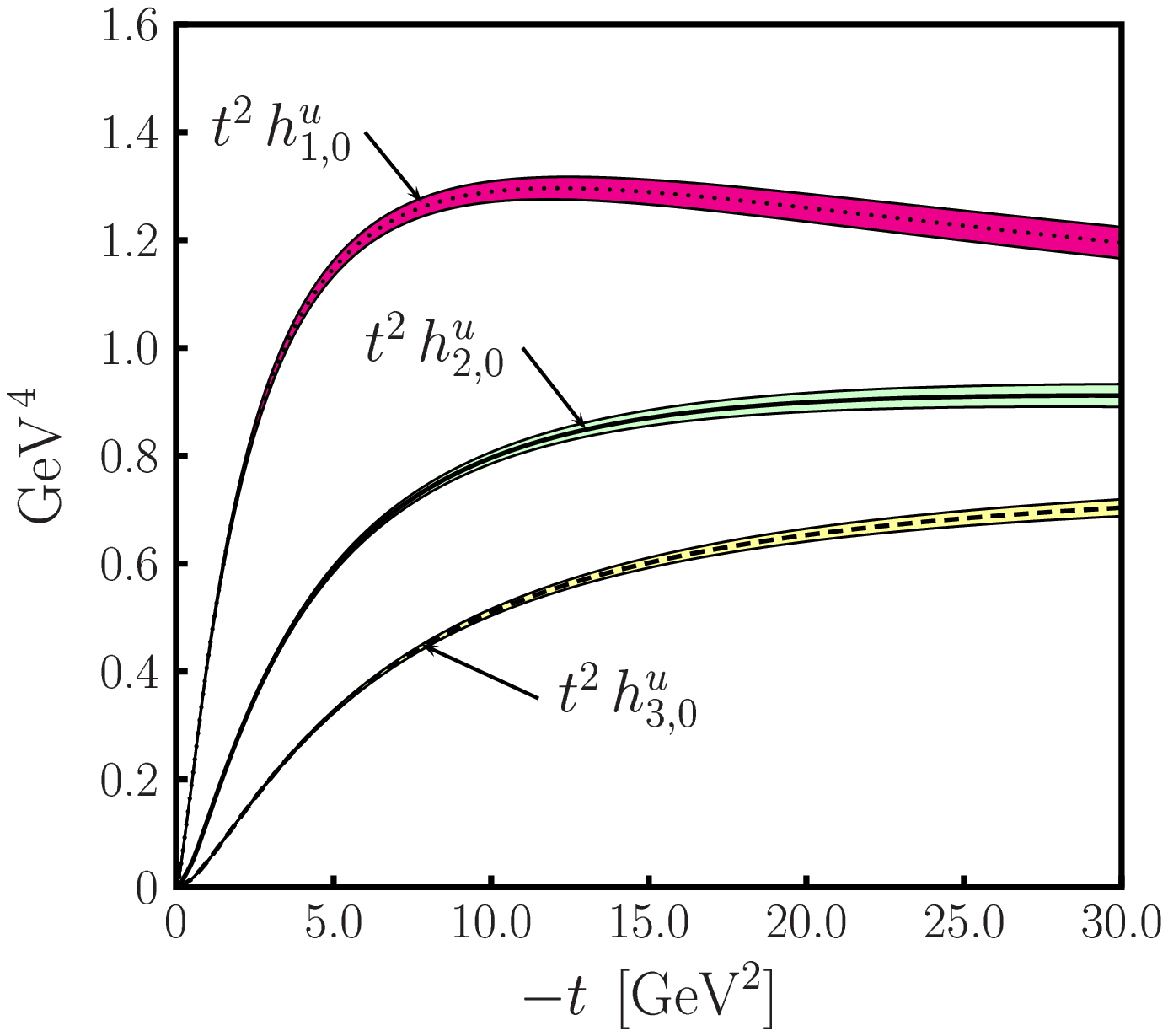}
\hspace{4em} 
\includegraphics[height=.35\textwidth,
 bb = 104 341 467 686,clip=true]{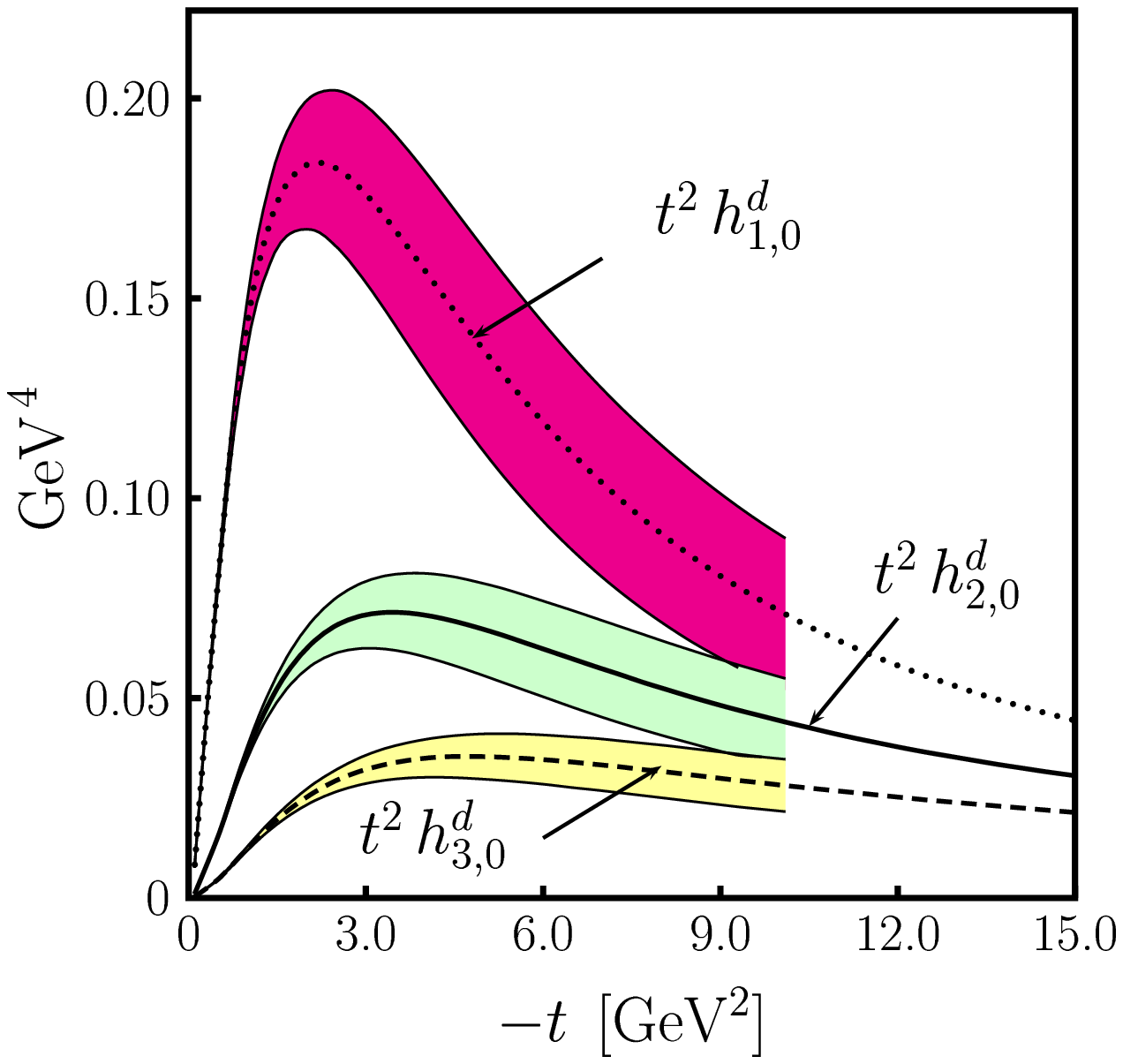}
\end{center}
\caption{\label{fig:Hmom} The first three moments of valence GPDs
$H_v^u$ (left) and $H_v^d$ (right), scaled with $t^2$.  The error
bands denote the parametric uncertainty resulting from the fit to the
Dirac form factors $F_1^{p}$ and~$F_1^{n}$.}
\end{figure}

The lowest moments of the GPD $H$  defined by
\be
h_{n,0}^q(t)\= \int_0^1 dx\, x^{n-1}\, H^q_v(x,t)\,,
\ee 
are shown in Fig.\ \ref{fig:Hmom}. The different powers with which the $u$
und $d$-quark moments drop correspond to the large-$x$ behaviour of
the parton distributions as I discussed in Sect.\ \ref{sec:comments},
see \req{powers}. Note also the large difference in magnitude between
the $u$ and $d$ moments. Strengthened by the charge factor the
$u$-quark contribution dominates the proton form factor in the $\gev$
region, the $d$-quark contribution amounts to less than $10\%$. Its
contribution to the neutron form factor is about $30\%$. High quality
neutron form factor data above $3\,\gev^2$ would allow for a direct
examination of the different powers.     

In contrast to the usual parton distribution which only provide
information on the longitudinal distribution of quarks inside the
nucleon, GPDs also give access to the transverse structure of the
nucleon by Fourier transforming the GPD with respect to
$\sqrt{-t}$. This is discussed in some detail in \cite{DFJK4}. One can
now also evaluate the valence quark contribution to the orbital
angular momentum the quarks inside the proton carry by exploiting Ji's
sum rule \cite{ji97}. A value of -0.17 has been found in \cite{DFJK4}
for the average valence quark contribution to the orbital angular momentum.
\section{Results for Compton scattering}
\label{sec:Compton}
\begin{figure}
\begin{center}
\includegraphics[width=.35\textwidth, height=.35\textwidth,
  bb= 50 107 387 430,clip=true]{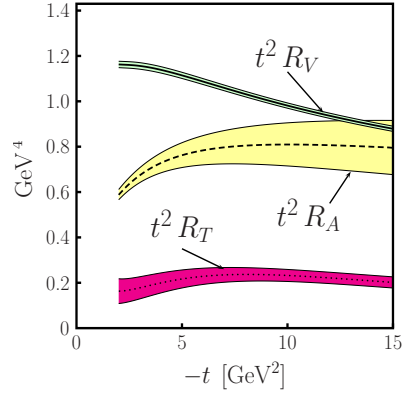}
\end{center}
\caption{\label{fig:comptonff} Results for the scaled Compton form
  factors obtained in \protect\cite{DFJK4}.}
\end{figure}

With the results for the GPDs at hand one can now evaluate the Compton form
factors 
\ba
R_V(t) &\simeq&\sum_q e_q^2 \int_{0}^1 \frac{dx}{x}\, H^q_v(x,t)\,, \nn\\
R_A(t) &\simeq& \sum_q e_q^2 \int_{0}^1 \frac{dx}{x}\, 
\widetilde{H}^q_v(x,t)\,, \nn\\
R_T(t) &\simeq&\sum_q e_q^2 \int_{0}^1 \frac{dx}{x}\, E^q_v(x,t)\,,
\label{Compton-formfactors}
\ea
Contributions from sea quarks are neglected. Numerical results for the Compton
form factors are shown in Fig.\ \ref{fig:comptonff}. 
Approximately the form factors $R_i$ behave $\sim t^{-2}$.
The particular flat behaviour of the scaled form factor $t^2R_T$ is a
consequence of a cancellation between the $u$ and $d$-quark
contributions. The result for $R_T$ is however subject to rather large
uncertainties, given the considerable freedom one encounters in
extracting $E$ from the Pauli form factor alone. The ratio $R_T/R_V$
behaves differently from the corresponding ratio of their
electromagnetic analogues $F_2$ and $F_1$. 

\begin{figure}
\begin{center}
\includegraphics[width=.35\textwidth, height=.35\textwidth,
  bb= 138 223 480 556,clip=true]{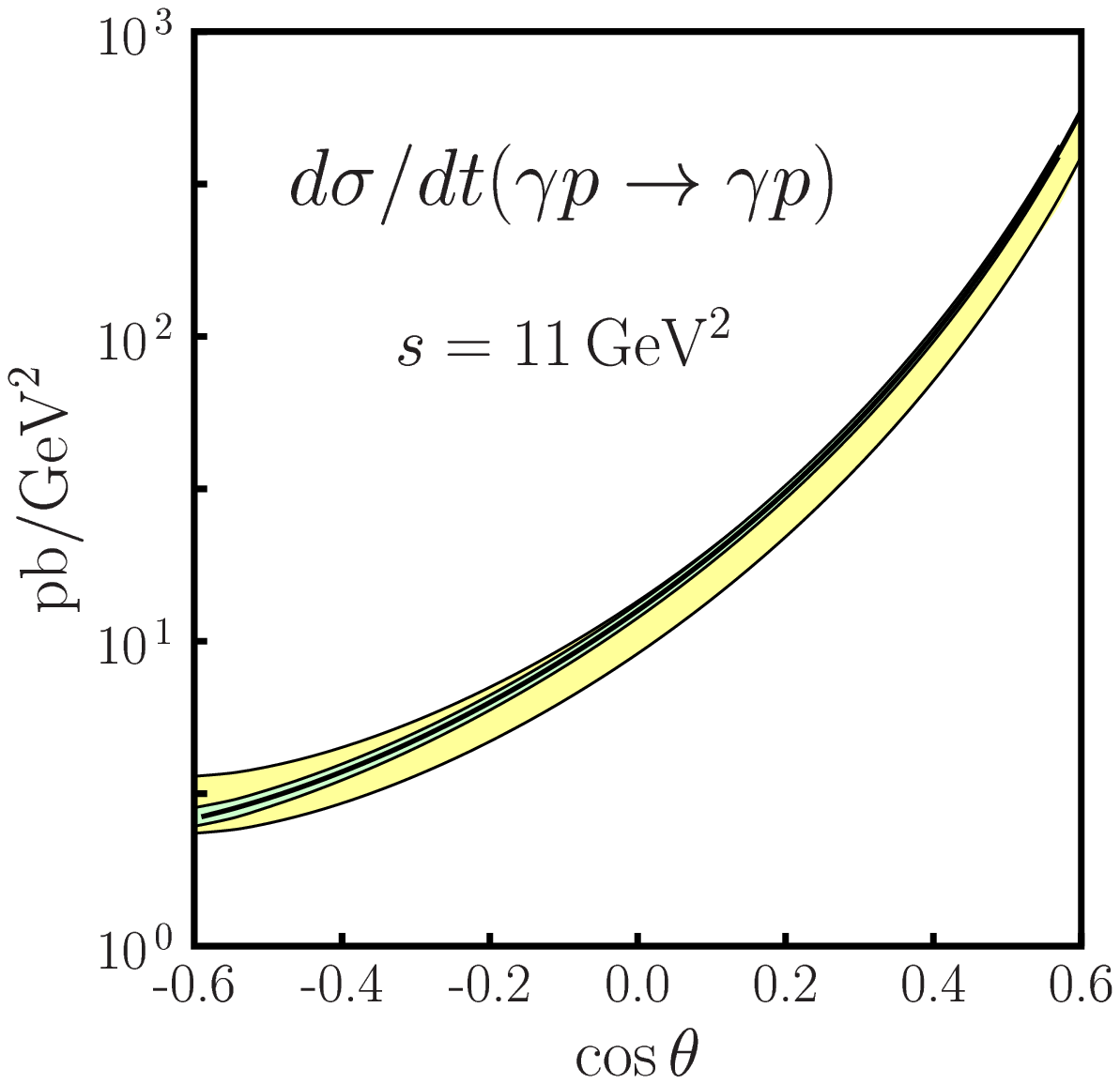}
\hspace{4em}
\includegraphics[width=.35\textwidth, height=.35\textwidth,
  bb= 122 340 433 684,clip=true]{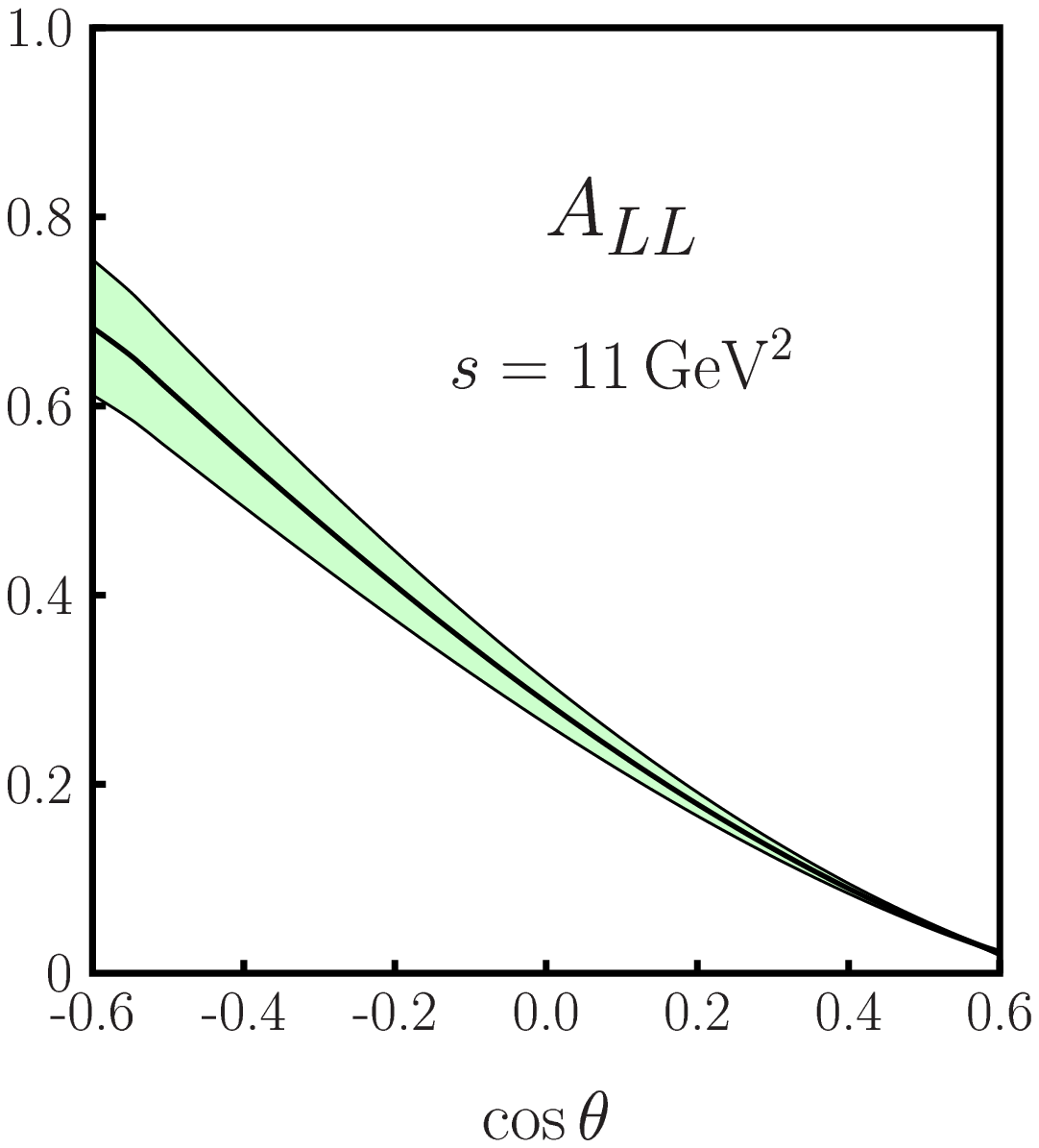}
\end{center}
\caption{\label{fig:wacs-obs} The unpolarized cross section (left) and
the helicity correlation parameter $A_{LL}$ (right) for wide-angle
Compton scattering at $s= 11 \gev^2$ as a function of the scattering
angle $\theta$ in the c.m.  Both observables are evaluated at NLO with
the Compton form factors shown in Fig.~\protect\ref{fig:comptonff}.}
\end{figure}

Inserting these Compton form factors into Eqs.\ \req{dsdt} and
\req{all}, one is now able to predict the Compton cross section in the
wide-angle region as well as the helicity correlation $A_{LL}=K_{LL}$. The
results for $s=11\,\gev^2$ are shown in Fig.\ \ref{fig:wacs-obs}. The
inner band for the curve of $d\sigma/dt$ reflects the parametric
errors of the form factors, essentially that of the vector form factor
which dominates the cross section. The other form factors contribute
less than $10\%$. The outer band indicates an estimate of the target mass
corrections, see \cite{DFHK}. The prediction for the cross section
from the handbag approach are substantially larger than those from a 
leading-twist calculation \cite{dixon}. The JLab E99-114 collaboration~\cite{nathan}
will provide accurate cross section data soon which will allow for a
crucial examination of the predictions from the handbag mechanism.

The JLab E99-114 collaboration \cite{nathan} has presented a first
measurement of $K_{LL}$ at a c.m.s.\ scattering angle of
$120^\circ$ and a photon energy of $3.23 \gev$. This still
preliminary data point is in fair agreement with the predictions
from the handbag given the small energy at which they are
available. The kinematical requirement of the handbag mechanism 
$s,\; -t,\; -u \gg \Lambda^2$ is not well satisfied and therefore one
has to be aware of large dynamical and kinematical corrections. 
It is to be stressed that the leading-twist 
approach~\cite{dixon} leads to a negative value for $K_{LL}$ at
angles larger than $90^\circ$ in conflict with the JLab 
result~\cite{nathan}. 

The handbag approach also applies to wide-angle photo- and
electroproduction of pseudoscalar and vector mesons.
The amplitudes again factorize into a parton-level subprocess, $\gamma
q\to M q$, and form
factors which represent $1/x$-moments of GPDs \cite{hanwen}. Their
flavor decomposition differs from those appearing in Compton
scattering, see \req{Compton-formfactors}. Here, it reflects the valence
quark structure of the produced meson. Since the GPDs and, hence, the
form factors for a given flavor, $R_i^q$, $i=V,A,T$ are process
independent they are known from the analysis of Ref.~\cite{DFJK4} for
$u$ and $d$ quarks (if the contribution from sea quarks can be ignored). 
Therefore, the soft physics input to calculations of photo-and
electroproduction of pions and $\rho$ mesons within the handbag
approach is now known.
\section{Summary}
The treatment of wide-angle exclusive reactions is not simple within
QCD and careful analyses are required. While the dominance of the 
leading-twist contribution for which all valence quarks of the
involved hadrons participate in the partonic
subprocess is expected for asymptotically large momentum transfer, it
seems that the handbag mechanism, characterized by one active parton,
dominates for momentum transfer of the order of $10\,\gev^2$. The
approximate power behaviour seen in many exclusive observables cannot
safely be considered as evidence for the dominance of the
leading-twist contribution. A careful investigation of the
perturbatives logs as well as an understanding of the normalization is
required. Soft mechanisms like the handbag can also explain the power
behaviour. There are many interesting predictions from the handbag
mechanism, some are in fair agreement with experiment, others still
awaiting their experimental examination. 

In Ref.~\cite{DFJK4} a first analysis of the GPDs has been performed
in analogy to those of the usual parton distribution, see for instance
\cite{Cteq}. A physically motivated parameterization of the GPDs have
been fitted to the available data on the form factors of the
nucleon. This analysis therefore provides information only on the GPDs
at zero skewness and for valence quarks. This suffices to evaluate
the soft physics input to wide-angle exclusive reactions but not for
deep virtual exclusive processes where the GPDs at non-zero skewness
are required. For these processes we still have to rely on models, see
for instance the review \cite{goeke}.


\end{document}